\pdfoutput=1
\documentclass[11pt]{article}

\usepackage{amsmath,amssymb}
\usepackage{amscd,amsthm}
\usepackage{amscd}

\usepackage{multirow}
\usepackage{accents}
\usepackage{hyperref}
\usepackage{color}
\usepackage[usenames,dvipsnames]{xcolor}
\usepackage{graphicx}
\usepackage{enumitem}

\usepackage{algorithm}
\usepackage{algorithmic}

\usepackage{setspace}

\usepackage{graphicx} 

\usepackage[
pass,
]{geometry}

\newgeometry{top=1in, bottom=1in, left=1in, right=1in}

\newfont{\mycrnotice}{ptmr8t at 7pt}
\newfont{\myconfname}{ptmri8t at 7pt}

\title{Practice of Efficient Data Collection via Crowdsourcing at Large-Scale}
\date{August 08, 2019}

\author{
Alexey	Drutsa, 
Viktoriya Farafonova, Valentina Fedorova, 
\\
Olga Megorskaya, Evfrosiniya Zerminova, Olga Zhilinskaya\thanks{Affiliation: Yandex (16, Leo Tolstoy St., Moscow, Russia, 119021; www.yandex.com).}
}

\begin{document}

\maketitle

\begin{abstract}
Modern machine learning algorithms need large datasets to be trained.
Crowdsourcing has become a popular approach to label large datasets in a shorter time as well as at a lower cost comparing to that needed for a limited number of experts. However, as crowdsourcing performers are non-professional and vary in levels of expertise, such labels are much noisier than those obtained from experts. For this reason, in order to collect good quality data within a limited budget special techniques such as incremental relabelling, aggregation and pricing need to be used.
We make an introduction to data labeling via public crowdsourcing marketplaces and present key components of efficient label collection. We show how to choose one of real label collection tasks, experiment with selecting settings for the labelling process, and launch label collection project at Yandex.Toloka, one of the largest crowdsourcing marketplace. The projects will be run on real crowds.
We also present main algorithms for aggregation, incremental relabelling, and pricing in crowdsourcing. In particular, we, first, discuss how to connect these three components to build an efficient label collection process; and, second, share rich industrial experiences of applying these algorithms and constructing large-scale label collection pipelines (emphasizing best practices and common pitfalls). 
\end{abstract}

\section{Introduction}
Modern machine learning algorithms require a large amount of labelled data to be trained. Crowdsourcing has become a popular source of such data due to its lower cost, higher speed, and diversity of opinions comparing to labelling data with experts. However, performers at crowdsourcing marketplaces are non-professional and their labels are much noisier than that of experts~\cite{snow/etal:2008}. For this reason, to obtain good quality labels via crowdsourcing and under a limited budget, special methods for label collection and processing are needed. The goal of this tutorial is to teach participants how to efficiently use crowdsourcing marketplaces for labelling data. 

\begin{figure}
	\centering
	\includegraphics[width=\textwidth]{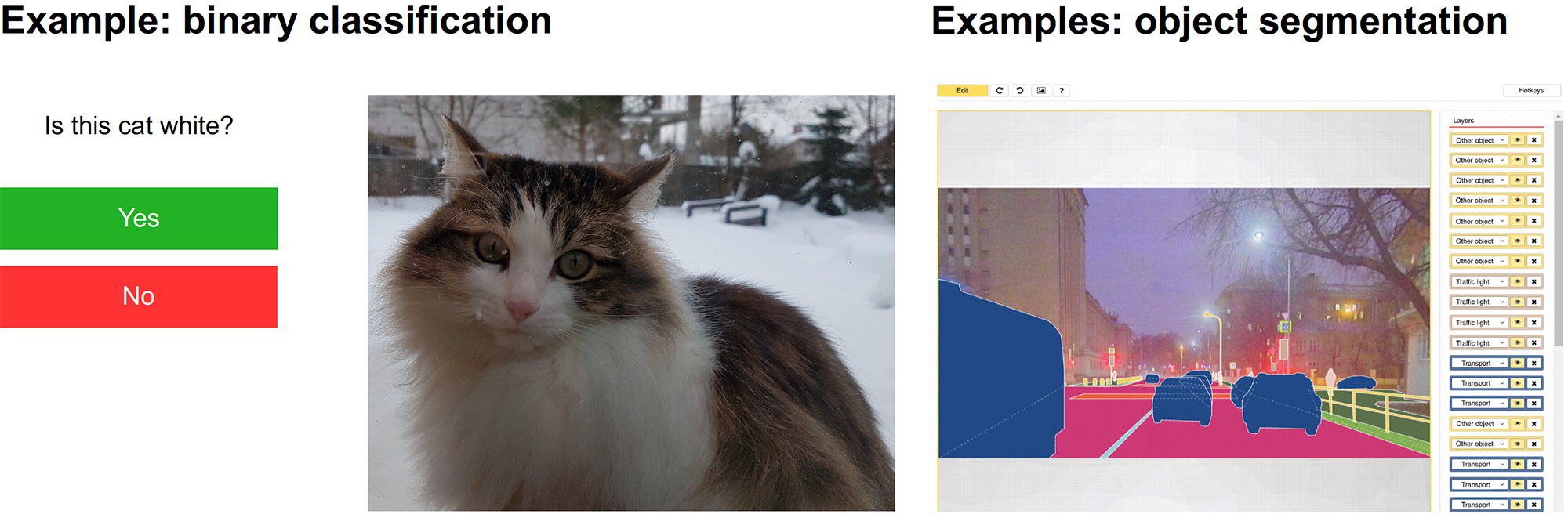}
	\vspace{-0.8cm}
	\caption{Examples of human intelligence tasks (HITs) that can be executed on crowdsourcing platforms: binary classification (the left side) and object segmentation (the right side).}
	\label{examples}
\end{figure}

Crowdsourcing platforms can process a wide range of tasks (a.k.a., human intelligence tasks, HITs), for instance:
information assessment (e.g., used in ranking of search results);
content categorization (e.g., used in text and media moderation, data cleaning and filtering);
content annotation (e.g., used in metadata tagging);
pairwise comparison (e.g., used in offline evaluation, media duplication check);
object segmentation, including 3D (e.g., used in image recognition for self-driving car);
audio and video transcription (e.g., used in speech recognition for voice-controlled virtual assistant);
field surveys (e.g., used to verify business information and office hours); etc. Two examples of tasks are in Figure~\ref{examples}.

\begin{figure}
	\centering
	\includegraphics[width=\textwidth]{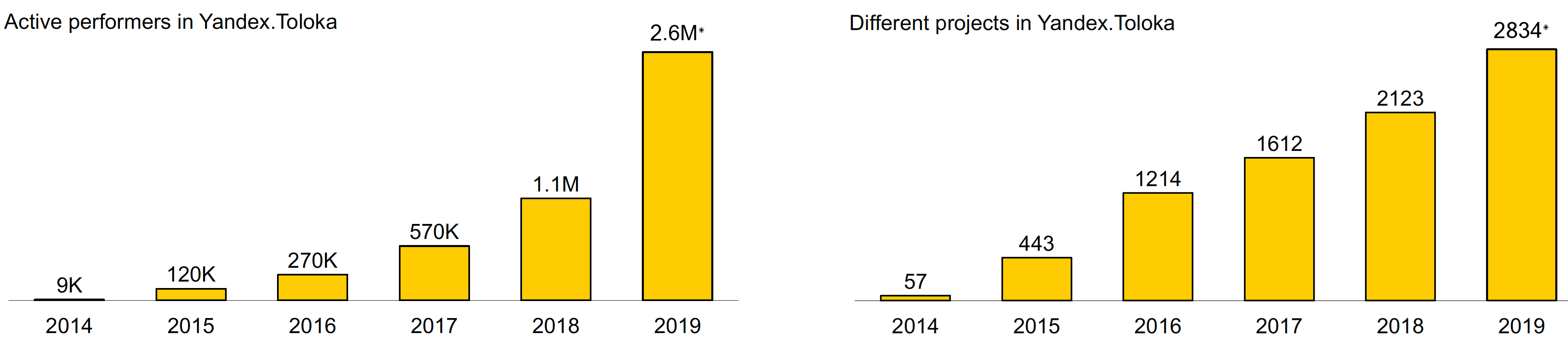}
		\vspace{-0.8cm}
	\caption{Crowdsourcing growth: Yandex experience (* statistic for 2019 is obtained via an extrapolation based on the first 7 months of 2019).}
	\label{growth}
\end{figure}

Crowdsourcing is widely used in modern industrial IT companies in permanent manner and on a large scale. The development of their products and services strongly depends on the quality and costs of labelled data. 
For instance, Yandex's crowdsourcing experience is presented in Figure~\ref{growth}, where the substantial growth is seen in terms of both active performers and projects. Currently,  25K performers execute around 6M HITs in more than 500 different projects everyday at Yandex.Toloka\footnote{\url{https://toloka.yandex.com/for-requesters}}.

\section{Key components for efficient data collection}
We discuss \emph{key components} required to collect labelled data: 
proper decomposition of tasks (construction of a pipeline of several small tasks instead of one large human intelligent task),
easy to read and follow task instructions,
easy to use task interfaces,
quality control techniques,
an overview of aggregation methods, and pricing.
Quality control techniques include: 
approaches ``before" task execution (selection of executors, education and exam tasks),
the ones ``within" task execution (golden sets, motivation of performers, tricks to remove bots and cheaters), and
approaches ``after" task execution (post verification/acceptance, consensus between performers).
We share best practices, including: pitfalls when designing instructions \& interfaces, important settings in different types of templates, training and examination for performers selection, important aspects in tasks instructions for performers, pipelines for evaluating the process of labelling.

\begin{figure}
	\centering
	\includegraphics[width=\textwidth]{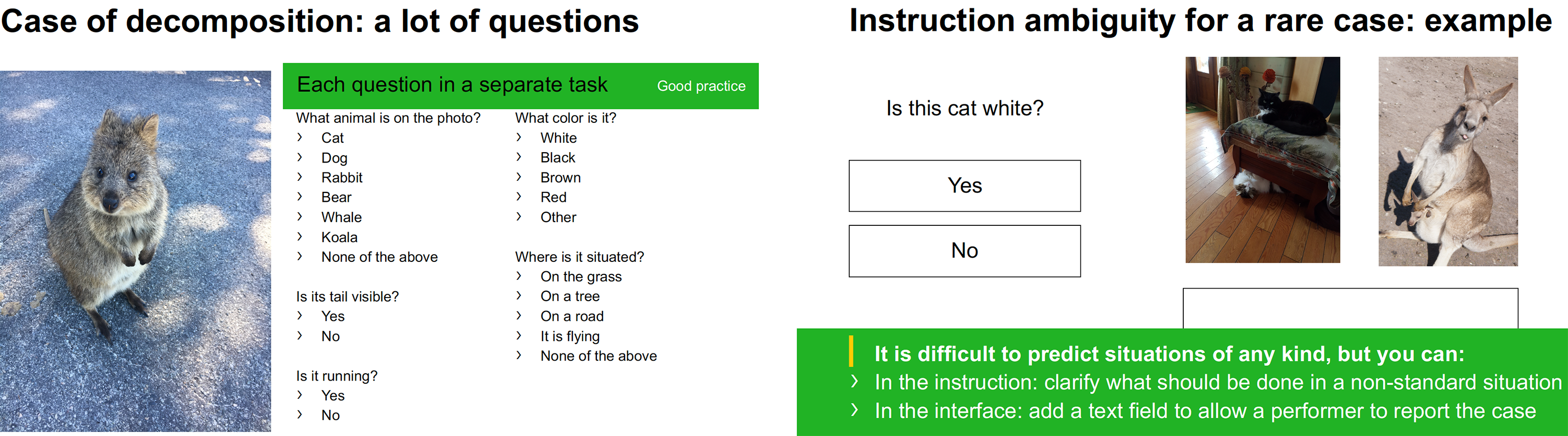}
		\vspace{-0.8cm}
	\caption{The left side: an example of a single task with multiple questions. The right side: an example of rare cases that should be taken into account when building task interfaces and instructions. }
	\label{example_decomp_instruct}
\end{figure}

Figure~\ref{example_decomp_instruct} (left side) contains an example of a single task with multiple questions. In this case, the best practice is to split this task into several one such that each question will be in a separate HIT.
Figure~\ref{example_decomp_instruct} (right side) contains few example images for a binary classification task where a performer should decide whether a cat is white or not. These examples are rare cases that should be taken into account when building task interfaces and instructions.

\begin{figure}
	\centering
	\includegraphics[width=\textwidth]{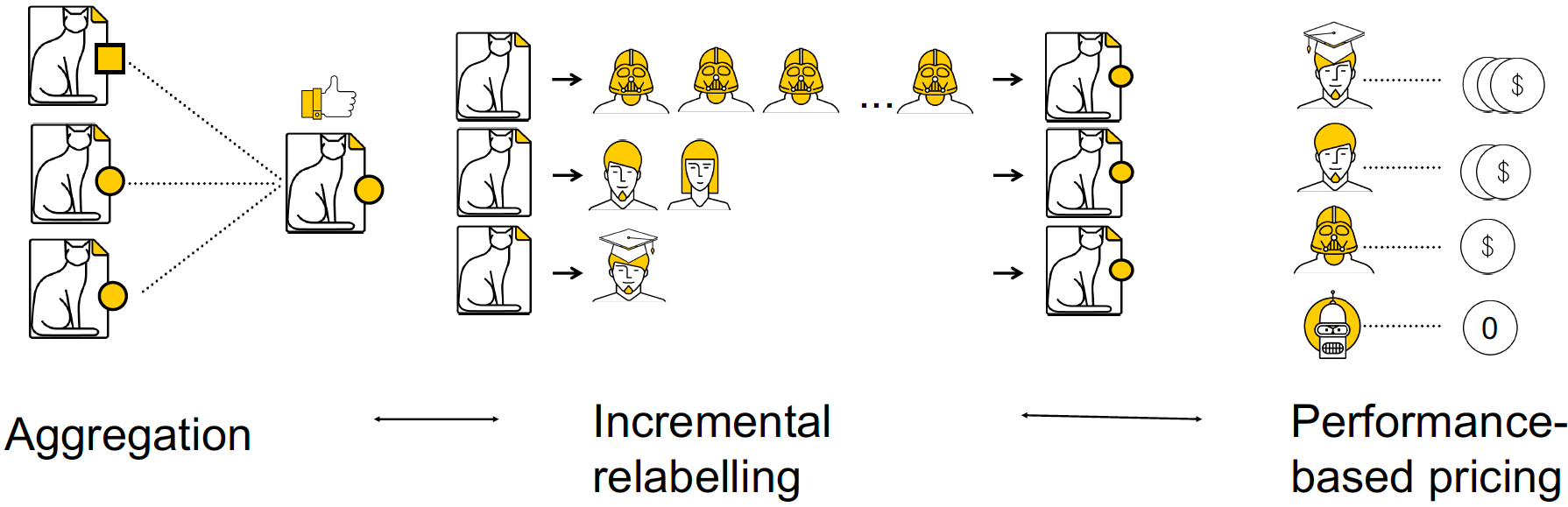}
	\vspace{-0.8cm}
	\caption{Interconnection between three approaches  that  make  crowdsourcing more efficient: aggregation, incremental relabelling (IRL),  and performance-based pricing.}
	\label{agg_IRL_PBP}
\end{figure}

\section{Efficiency methods: aggregation, IRL, and pricing}
The next approaches are the main ones that  make  crowdsourcing more efficient:
\begin{itemize}
	\item \emph{Methods for aggregation in crowdsourcing}. Classical models: Majority Vote, Dawid-Skene~\cite{dawid/skene:1979}, GLAD~\cite{whitehill/etal:2009}, Minimax Entropy~\cite{zhou/etal:2015}. 
	Analysis of aggregation performance and difficulties in comparing aggregation models in unsupervised setting~\cite{sheshadri/lease:2013,imamura/etal:2018}.
	Advanced works on aggregation: combination of aggregation and learning a classifier~\cite{raykar/etal:2010}, using features of tasks and performers for aggregation~\cite{ruvolo/etal:2013,welinder/etal:2010,jin/etal:2017}, ensemble of aggregation models~\cite{faridani/buscher:2013}, aggregation of crowdsourced pairwise comparisons~\cite{chen/etal:2013}.
	\item \emph{Incremental relabelling (IRL)}. Motivation and the problem of incremental relabelling: IRL based on Majority Vote; IRL methods with performers quality scores~\cite{Ipeirotis:2014,ertekin/etal:2012,abraham/etal:2016}; active learning~\cite{Lin:2014}. Connections between aggregation and IRL algorithms. Experimental results of using IRL at crowdsourcing marketplaces.
	\item \emph{Pricing of tasks in crowdsourcing}. Practical approaches for task pricing~\cite{ho/etal:2015,wang/etal:2013,cheng/etal:2015,yin/etal:2013}. Theoretical background for pricing mechanisms in crowdsourcing: efficiency, stability, incentive compatibility, etc. Pricing experiments and industrial experience of using pricing at crowdsourcing platforms. 
\end{itemize}

\begin{figure}
	\centering
	\includegraphics[width=\textwidth]{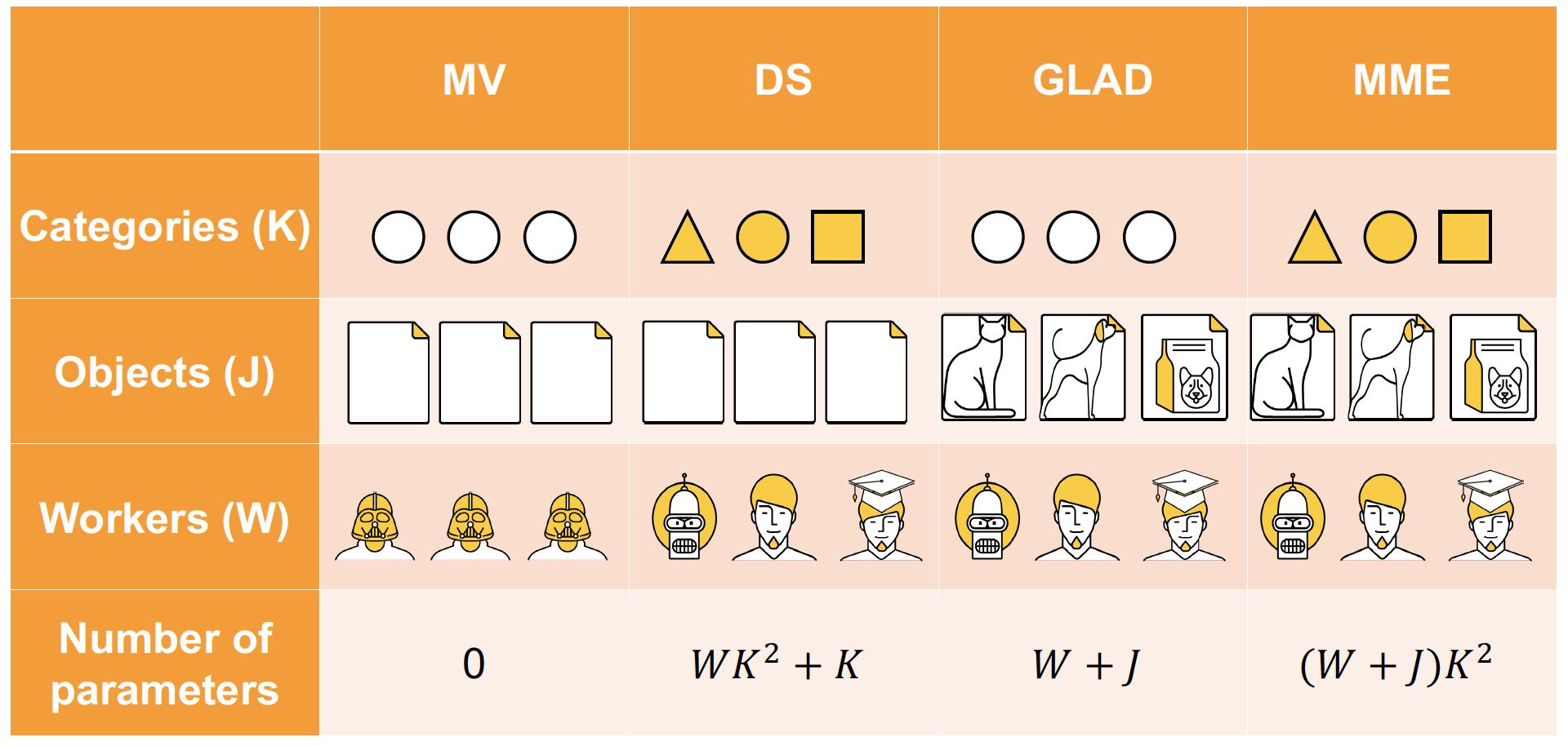}
		\vspace{-0.8cm}
	\caption{Summary on the key properties of the main aggregation methods: Majority Vote, Dawid-Skene~\cite{dawid/skene:1979}, GLAD~\cite{whitehill/etal:2009}, Minimax Entropy~\cite{zhou/etal:2015}.}
	\label{aggreg_compar}
\end{figure}

\section{Crowdsourcing pipeline to highlight objects on images}
\begin{figure}
	\centering
	\includegraphics[width=\textwidth]{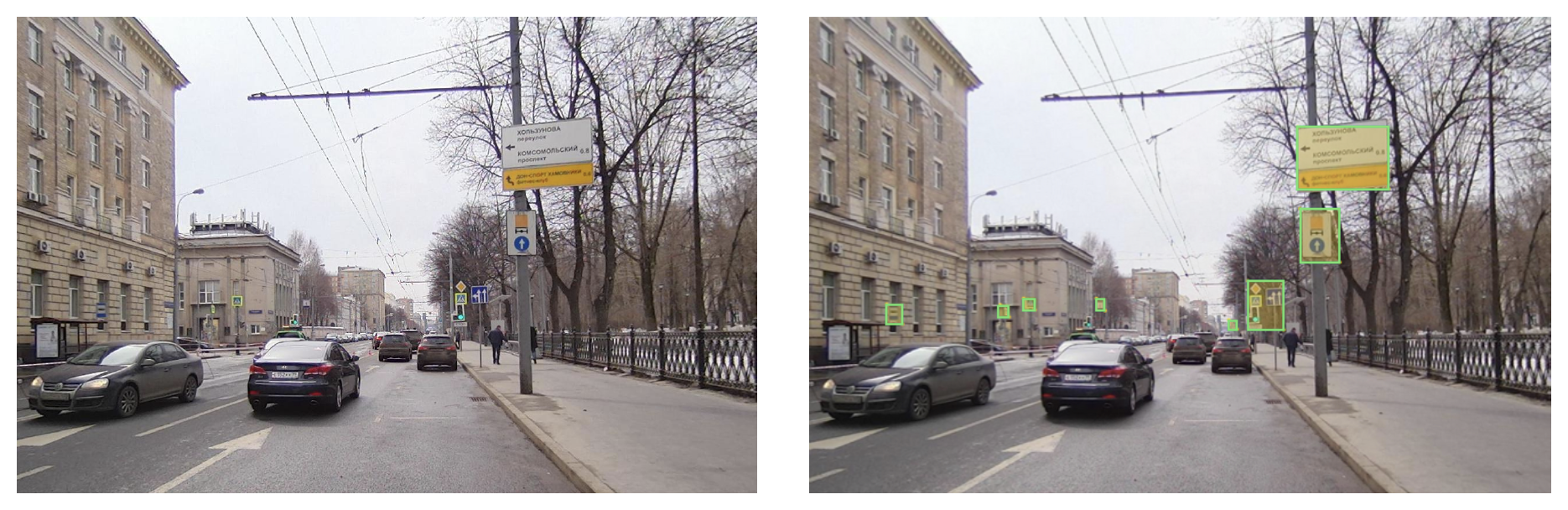}
	\vspace{-0.8cm}
	\caption{An example of a photo before (the left side) and after processing (the right side): all traffic signs are highlighted by bounding boxes.}
	\label{practice_before_after}
\end{figure}

\begin{figure}
	\centering
	\includegraphics[width=\textwidth]{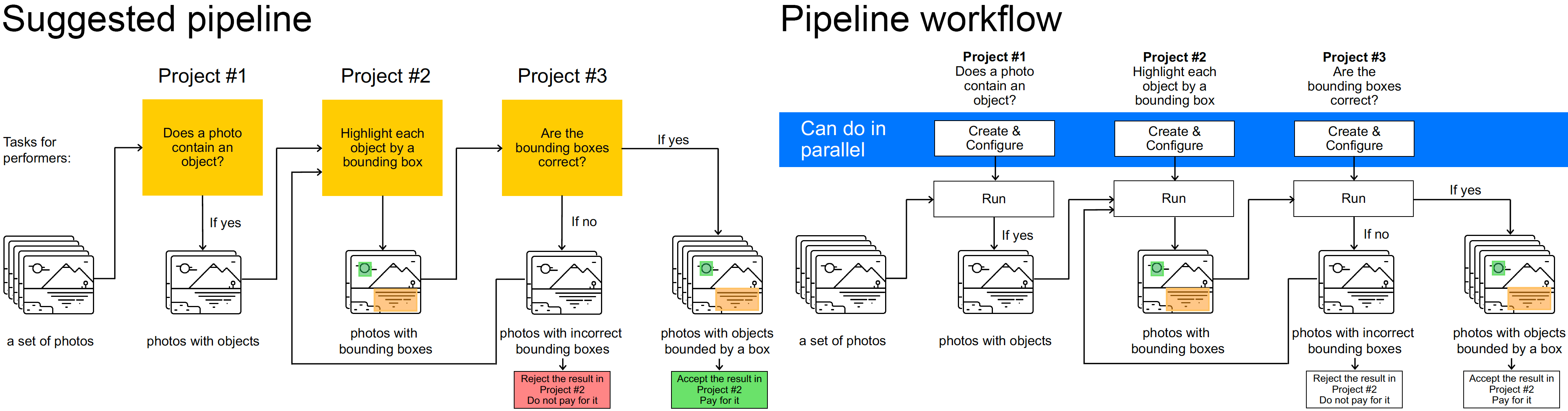}
	\vspace{-0.8cm}
	\caption{The left side: the suggested crowdsourcing pipeline to solve the problem of object highlighting on photos. The right side: how to work on creation and running of the suggested pipeline.}
	\label{practice_pipeline}
\end{figure}

Attendees  of our practice session create and run a crowdsourcing pipeline for a real problem on real performers. We propose to highlight objects of a certain type on images. A set of photos of real roads is taken as an example (since such a task is vital for autonomous vehicle development). 
Participants should select a type of objects to be highlighted: e.g., 
people,
transport,
road,
curb,
traffic lights,
traffic signs,
sidewalk,
pedestrian crossing, etc.
Highlighting of objects of the selected type is proposed to be done by means of bounding boxes via a public crowdsourcing platform.
The formal setup of our task is as follows: 
\begin{itemize}
\item each object of a selected type
\item in each photo from the dataset
\item needs to be highlighted by a rectangle (bounding box).
\end{itemize}
For instance, if traffic signs are chosen, then Figure~\ref{practice_before_after} demonstrates how a photo should be processed.
Participants propose their crowdsourcing pipelines and compare them with ours.
For the described task, we suggest to use the pipeline depicted in Figure~\ref{practice_pipeline} as the baseline.

\begin{figure}
	\centering
	\includegraphics[width=\textwidth]{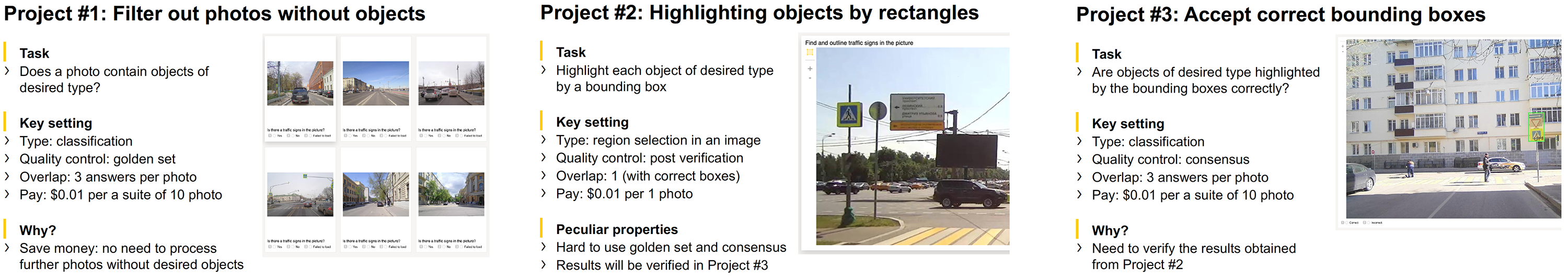}
	\vspace{-0.8cm}
	\caption{Short descriptions of HITs of three types used in the suggested crowdsourcing pipeline.}
	\label{practice_projects}
\end{figure}

This simple pipeline consists of three projects. 
The tasks for the first one are binary classification HITs. 
The second project contains HITs with a bounding box highlighting tool.
The third project designed to verify the results obtained from the second project.
The summary on these projects is shown in Figure~\ref{practice_projects}.

Attendees  of our practice session create, configure, and run this pipeline on real crowd performers.
We run this pipeline to process 100 images and highlight traffic signs. Our results are as follows.

\begin{enumerate}
	\item Project \#1: "Does a photo contain traffic signs?"
	\begin{itemize}
	\item 100 photos evaluated
	\item within 4 min on real performers
	\item cost: \$0.3 + Toloka fee
	\end{itemize}
	\item Project \#2: "Highlight each traffic sign by a bounding box"
\begin{itemize}
	\item 67 photos processed
	\item within 5.5 min on real performers
	\item cost: \$0.67 + Toloka fee
\end{itemize}
	\item Project \#3: "Are traffic signs highlighted by the bounding boxes correctly?"
\begin{itemize}
	\item 90 photos evaluated
	\item within 5 min on real performers
	\item cost: \$0.36 + Toloka fee
\end{itemize}
\end{enumerate}

\section{Related tutorials}
Previous tutorials consider different components of labeling process separately and did not include practice sessions. On the contrast, the goals of our tutorial is to explain the main algorithms for incremental relabelling, aggregation, and pricing and their connections to each other, and to teach participants the main principles for setting up an efficient process of labeling data at a crowdsourcing marketplace. Following is a summary of relevant topics covered in previous tutorials:
\begin{itemize}
	\item ``Crowdsourcing: Beyond Label Generation" presented at NIPS 2016, ALC 2017, and KDD 2017. A part of this tutorial devoted to an overview of empirical results about performers reaction to pricing.
	\item  ``Crowd-Powered Data Mining" conducted at KDD 2018. The introduction and the first part of this tutorial was devoted to the standard process of crowdsourcing label collection and aggregation.
	\item  ``Crowdsourced Data Management: Overview and Challenges" was held at SIGMOD'17 and partly focused on methods for aggregating crowdsourced data. 
	\item  ``Truth Discovery and Crowdsourcing Aggregation: A Unified Perspective" was conducted at VLDB 2015 and dedicated to methods for aggregating crowdsourced data. 
	\item  ``Spatial Crowdsourcing: Challenges, Techniques, and Applications" was conducted at VLDB 2016. This tutorial focused on using crowdsourcing for spatial tasks and included efficient methods for task targeting, aggregating data, and the effect of pricing for such tasks. Our tutorial will be devoted to another type of crowdsourcing tasks which is multiclassification.
\end{itemize}

\section*{Tutorial materials}
The tutorial materials (slides and instructions) are available at  \url{https://research.yandex.com/tutorials/crowd/kdd-2019}.

\nocite{dawid/skene:1979}
\nocite{whitehill/etal:2009}
\nocite{zhou/etal:2015}
\nocite{raykar/etal:2010}
\nocite{snow/etal:2008}
\nocite{ruvolo/etal:2013}
\nocite{faridani/buscher:2013}
\nocite{welinder/etal:2010}
\nocite{jin/etal:2017}
\nocite{imamura/etal:2018}
\nocite{sheshadri/lease:2013}
\nocite{kim/ghahramani:2012}
\nocite{venanzi/etal:2014}
\nocite{vuurens/etal:2011}
\nocite{chen/etal:2013}
\nocite{ipeirotis:2014}
\nocite{abraham/etal:2016}
\nocite{ertekin/etal:2012}
\nocite{Lin:2014}
\nocite{liu/wang:2012}
\nocite{zhao/etal:2011}
\nocite{wang/etal:2013}
\nocite{cheng/etal:2015}
\nocite{ho/etal:2015}
\nocite{difallah/etal:2014}
\nocite{yin/etal:2013}
\nocite{shah/etal:2015}
\nocite{shah/zhou:2016}

\bibliographystyle{plain}
\bibliography{2019-kdd-tutorial-extabstract}

\begin{thebibliography}{10}

\bibitem{abraham/etal:2016}
Ittai Abraham, Omar Alonso, Vasilis Kandylas, Rajesh Patel, Steven Shelford,
  and Aleksandrs Slivkins.
\newblock How many workers to ask?: Adaptive exploration for collecting high
  quality labels.
\newblock In {\em Proceedings of the 39th International ACM SIGIR conference on
  Research and Development in Information Retrieval}, pages 473--482, 2016.

\bibitem{chen/etal:2013}
X.~Chen, P.~N Bennett, K.~Collins-Thompson, and E.~Horvitz.
\newblock Pairwise ranking aggregation in a crowdsourced setting.
\newblock In {\em Proceedings of WSDM}, 2013.

\bibitem{cheng/etal:2015}
Justin Cheng, Jaime Teevan, and Michael~S Bernstein.
\newblock Measuring crowdsourcing effort with error-time curves.
\newblock In {\em Proceedings of the 33rd Annual ACM Conference on Human
  Factors in Computing Systems}, pages 1365--1374. ACM, 2015.

\bibitem{dawid/skene:1979}
A.~P. Dawid and A.~M Skene.
\newblock Maximum likelihood estimation of observer error-rates using the em
  algorithm.
\newblock {\em Applied statistics}, pages 20--28, 1979.

\bibitem{difallah/etal:2014}
Djellel~Eddine Difallah, Michele Catasta, Gianluca Demartini, and Philippe
  Cudr{\'e}-Mauroux.
\newblock Scaling-up the crowd: Micro-task pricing schemes for worker retention
  and latency improvement.
\newblock In {\em Second AAAI Conference on Human Computation and
  Crowdsourcing}, 2014.

\bibitem{ertekin/etal:2012}
Seyda Ertekin, Haym Hirsh, and Cynthia Rudin.
\newblock Learning to predict the wisdom of crowds.
\newblock {\em arXiv preprint arXiv:1204.3611}, 2012.

\bibitem{faridani/buscher:2013}
Siamak Faridani and Georg Buscher.
\newblock Labelboost: An ensemble model for ground truth inference using
  boosted trees.
\newblock In {\em First AAAI Conference on Human Computation and
  Crowdsourcing}, 2013.

\bibitem{ho/etal:2015}
Chien-Ju Ho, Aleksandrs Slivkins, Siddharth Suri, and Jennifer~Wortman Vaughan.
\newblock Incentivizing high quality crowdwork.
\newblock In {\em Proceedings of the 24th International Conference on World
  Wide Web}, pages 419--429. International World Wide Web Conferences Steering
  Committee, 2015.

\bibitem{imamura/etal:2018}
Hideaki Imamura, Issei Sato, and Masashi Sugiyama.
\newblock Analysis of minimax error rate for crowdsourcing and its application
  to worker clustering model.
\newblock {\em arXiv preprint arXiv:1802.04551}, 2018.

\bibitem{Ipeirotis:2014}
P~G Ipeirotis, F~Provost, V~S Sheng, and J~Wang.
\newblock Repeated labeling using multiple noisy labelers.
\newblock In {\em Data Mining and Knowledge Discovery}, pages 402--441.
  Springer, 2014.

\bibitem{jin/etal:2017}
Yuan Jin, Mark Carman, Dongwoo Kim, and Lexing Xie.
\newblock Leveraging side information to improve label quality control in
  crowd-sourcing.
\newblock In {\em Fifth AAAI Conference on Human Computation and
  Crowdsourcing}, 2017.

\bibitem{kim/ghahramani:2012}
H.~Kim and Z.~Ghahramani.
\newblock Bayesian classifier combination.
\newblock In {\em International conference on artificial intelligence and
  statistics}, pages 619--627, 2012.

\bibitem{Lin:2014}
Christopher~H Lin, M~Mausam, and Daniel~S Weld.
\newblock To re(label), or not to re(label).
\newblock In {\em Second AAAI conference on human computation and
  crowdsourcing}. AAAI, 2014.

\bibitem{liu/wang:2012}
Chao Liu and Yi-Min Wang.
\newblock Truelabel+ confusions: A spectrum of probabilistic models in
  analyzing multiple ratings.
\newblock {\em arXiv preprint arXiv:1206.4606}, 2012.

\bibitem{raykar/etal:2010}
V.~C Raykar, S.~Yu, L.~H Zhao, G.~H. Valadez, C.~Florin, L.~Bogoni, and L.~Moy.
\newblock Learning from crowds.
\newblock {\em The Journal of Machine Learning Research}, 11:1297--1322, 2010.

\bibitem{ruvolo/etal:2013}
P.~Ruvolo, J.~Whitehill, and J.~R Movellan.
\newblock Exploiting commonality and interaction effects in crowdsourcing tasks
  using latent factor models.
\newblock 2013.

\bibitem{shah/zhou:2016}
Nihar Shah and Dengyong Zhou.
\newblock No oops, you won’t do it again: Mechanisms for self-correction in
  crowdsourcing.
\newblock In {\em International conference on machine learning}, pages 1--10,
  2016.

\bibitem{shah/etal:2015}
Nihar Shah, Dengyong Zhou, and Yuval Peres.
\newblock Approval voting and incentives in crowdsourcing.
\newblock In {\em International Conference on Machine Learning}, pages 10--19,
  2015.

\bibitem{sheshadri/lease:2013}
Aashish Sheshadri and Matthew Lease.
\newblock Square: A benchmark for research on computing crowd consensus.
\newblock In {\em First AAAI conference on human computation and
  crowdsourcing}, 2013.

\bibitem{snow/etal:2008}
R.~Snow, B.~O'Connor, D.~Jurafsky, and A.~Y Ng.
\newblock Cheap and fast---but is it good?: evaluating non-expert annotations
  for natural language tasks.
\newblock In {\em Proceedings of the conference on empirical methods in natural
  language processing}, pages 254--263. Association for Computational
  Linguistics, 2008.

\bibitem{venanzi/etal:2014}
M.~Venanzi, J.~Guiver, G.~Kazai, P.~Kohli, and M.~Shokouhi.
\newblock Community-based bayesian aggregation models for crowdsourcing.
\newblock In {\em Proceedings of the 23rd international conference on World
  wide web}, pages 155--164, 2014.

\bibitem{vuurens/etal:2011}
Jeroen Vuurens, Arjen~P de~Vries, and Carsten Eickhoff.
\newblock How much spam can you take? an analysis of crowdsourcing results to
  increase accuracy.
\newblock In {\em Proc. ACM SIGIR Workshop on Crowdsourcing for Information
  Retrieval (CIR'11)}, pages 21--26, 2011.

\bibitem{wang/etal:2013}
Jing Wang, Panagiotis~G Ipeirotis, and Foster Provost.
\newblock Quality-based pricing for crowdsourced workers.
\newblock 2013.

\bibitem{welinder/etal:2010}
Peter Welinder, Steve Branson, Pietro Perona, and Serge~J Belongie.
\newblock The multidimensional wisdom of crowds.
\newblock In {\em Advances in neural information processing systems}, pages
  2424--2432, 2010.

\bibitem{whitehill/etal:2009}
J.~Whitehill, T.~Wu, J.~Bergsma, J.~R Movellan, and P.~L Ruvolo.
\newblock Whose vote should count more: Optimal integration of labels from
  labelers of unknown expertise.
\newblock In {\em Advances in neural information processing systems}, pages
  2035--2043, 2009.

\bibitem{yin/etal:2013}
Ming Yin, Yiling Chen, and Yu-An Sun.
\newblock The effects of performance-contingent financial incentives in online
  labor markets.
\newblock In {\em Twenty-Seventh AAAI Conference on Artificial Intelligence},
  2013.

\bibitem{zhao/etal:2011}
Liyue Zhao, Gita Sukthankar, and Rahul Sukthankar.
\newblock Incremental relabeling for active learning with noisy crowdsourced
  annotations.
\newblock In {\em 2011 IEEE Third International Conference on Privacy,
  Security, Risk and Trust and 2011 IEEE Third International Conference on
  Social Computing}, pages 728--733. IEEE, 2011.

\bibitem{zhou/etal:2015}
D.~Zhou, Q.~Liu, J.~C Platt, C.~Meek, and N.~B Shah.
\newblock Regularized minimax conditional entropy for crowdsourcing.
\newblock {\em arXiv preprint arXiv:1503.07240}, 2015.

\end{thebibliography}

\appendix

\section{Introduction to the requester interface}
\emph{Interface for requesters} is discussed on the example of the  crowdsourcing marketplace  Yandex.Toloka. This include key concepts and definitions: projects and task instructions, templates for projects, pools of tasks, task suites, honeypots, quality control, performer skills, tasks with post acceptance and auto acceptance. Project creation includes quick start and main settings for labelling data.
Types of project templates are multiple choice task to classify items, side by side comparisons, surveys to collect opinions on a certain topic, audio transciption, voice recording, object selection to locate one or more objects in an image, spatial task to visit a certain place and perform  a simple activity.

Key types of instances in Yandex.Toloka are  a project, a pool, and a task. 
A project defines the structure of tasks and how to perform them. A requester configures a task interface,  a task instruction, input and output data types in a project.
A pool is a batch of tasks and defines access of performers. A requester configures performer filters, quality control mechanisms,  overlap settings, and pricing in a pool.
A task is a particular input data and results for it from performers.

\begin{figure}
	\centering
	\includegraphics[width=\textwidth]{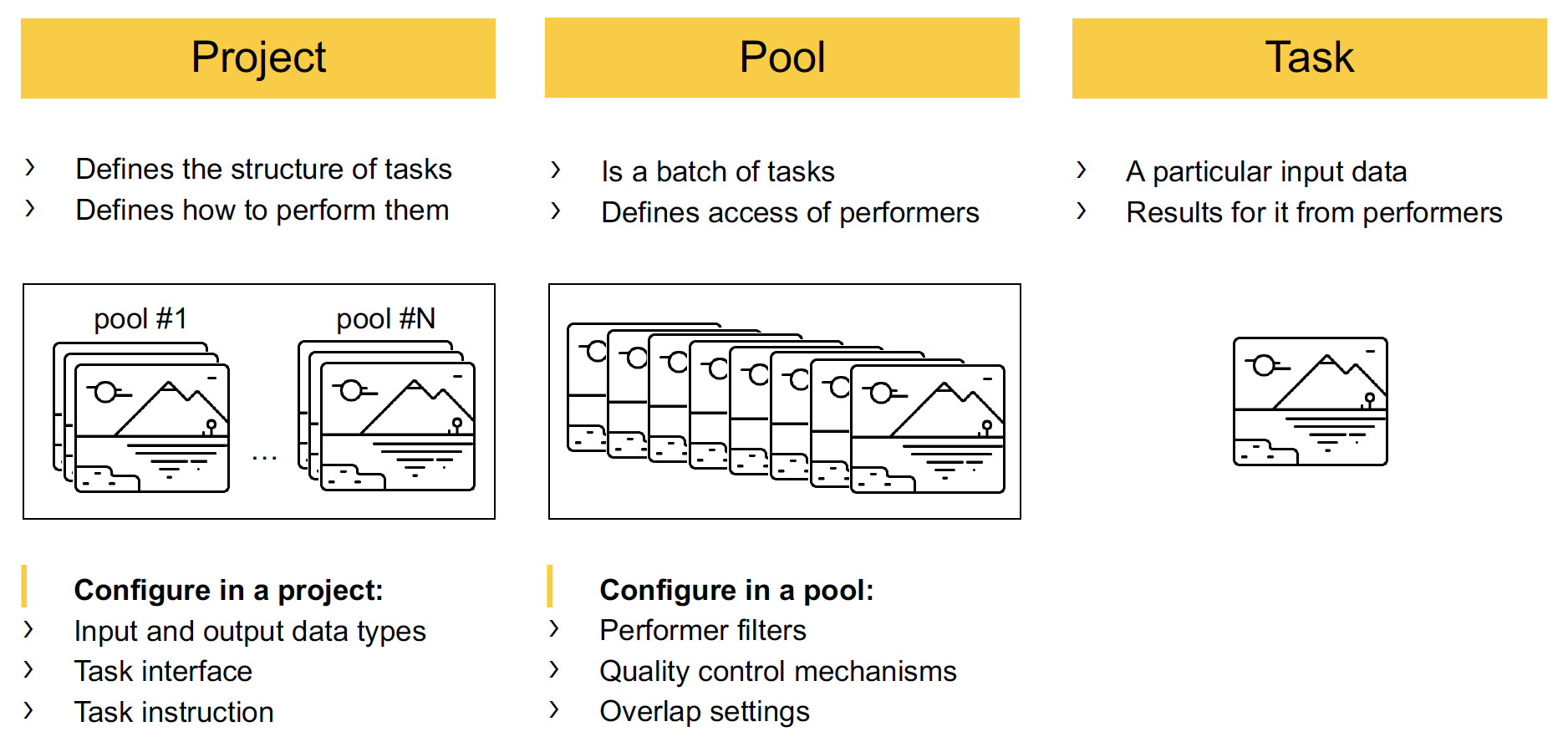}
		\vspace{-0.8cm}
	\caption{Key types of instances in Yandex.Toloka: a project, a pool, and a task.}
	\label{toloka_instances}
\end{figure}

\end{document}